# Reservoir Computing with Heterogeneous Magnetic Metamaterials

R. Yagan[1], C. Swindells[2], I. T. Vidamour[3], G. Venkat[4], J. Griffiths[3], E. Vasilaki[3], M. O. A. Ellis[3] and T. J. Hayward[1]

[1]School of Chemical, Material and Biological Engineering, The University of Sheffield, Sheffield, United Kingdom
[2]National Institute of Standards and Technology, Boulder, CO, United States
[3]School of Computer Science, The University of Sheffield, Sheffield, United Kingdom
[4]Diamond Light Source, Harwell Science and Innovation Campus, Didcot, Oxfordshire, United Kingdom

**Abstract**

Physical reservoir computing utilizes the intrinsic nonlinear and history-dependent dynamics of physical systems to perform machine-learning tasks with minimal training overhead. Here, we introduce a nanomagnetic reservoir computer based on a heterogeneous array of interconnected magnetic nanorings, combined with multi-channel planar Hall effect readout. The device comprises subarrays of rings with systematically varied track widths ranging from 500 nm to 300 nm, enabling access to the heterogeneous dynamics of geometrically diverse magnetic systems within a single reservoir. By applying time-varying input signals as modulations of a driving rotating magnetic field, we evaluate the nanoring reservoir's performance on nonlinear signal transformation and Mackey-Glass time-series prediction tasks. We find that combining outputs from multiple width-dependent channels significantly reduces the normalized root-mean-square error compared to single-channel readout, with the optimal channel combinations depending on task requirements. These results demonstrate that geometric heterogeneity provides an additional, experimentally accessible degree of freedom and complementary computational features. Principal component analysis further reveals that a reduced subset of correlated features captures most of the computationally relevant information while suppressing noise contributions. These results demonstrate that controlled geometric heterogeneity enhances reservoir expressivity and suggest a route toward scalable magnetic computing architectures in which multi-output magnetic metamaterials serve as configurable dynamical building blocks for device networks.

---

Reservoir computing (RC) is a computational paradigm originally developed for recurrent neural networks such as echo state networks[1] and liquid state machines[2], and is particularly well-suited for processing temporal and sequential data due to its ability to map time-varying inputs into a high-dimensional dynamic state space while requiring only a linear readout layer for training.[3] This simplicity of training reservoir computers, by eliminating the high computational cost and instability associated with back-propagation through time,[4] makes it attractive for application to simple, nonlinear signal transformation (SigT), prediction, and classification tasks, with well-known benchmarks, such as Mackey–Glass time-series prediction (MG-TSP)[5] tasks, where reservoirs can outperform traditional models with orders of magnitude fewer trained parameters. Physical RC extends this concept by using the intrinsic dynamics of physical substrates as the reservoir, enabling computation 'in materia' with minimal external control and low training overhead. The literature of physical RC highlights the promise of this approach for energy-efficient, high-speed computing architectures that are attractive for edge and embedded applications, since physical systems can naturally exhibit the nonlinearity, memory, and rich state spaces that underpin reservoir dynamics.[6–8] A physical RC may be realized using a diverse range of physical phenomena, including photonic delay systems,[9] mechanical vibrations,[10] memristive networks,[11] and, specifically, magnetic and spintronic devices,[12] whose magnetization dynamics inherently provide fading memory and nonlinearity without the need for adaptive training of internal weights.

RC has been explored using a variety of magnetic and spintronic systems spanning different dynamical regimes, including spin waves, spin-torque oscillators, magnetic tunnel junctions, domain walls, and skyrmions.[8,13–16] At short timescales (MHz-GHz), magnetic reservoirs exploit nonlinear magnetization dynamics governed by the Landau-Lifshitz-Gilbert equation, including precession, damping, and spin-torque-driven excitations.[17] These can be found in spin-torque oscillators,[13] magnetic tunnel junctions,[14] and spin waves[18] in magnetic media. At longer timescales, computation can instead emerge from thermally activated dynamics or field-driven switching between non-volatile magnetic states, such as domain-wall networks[16] and artificial spin systems[19], providing nonlinear responses and fading memory through stochastic or hysteretic evolution.[20]

In our previous work, we demonstrated that interconnected magnetic nanoring arrays driven by rotating magnetic fields provide a robust physical reservoir, where rich emergent dynamics arise from domain-wall (DW) nucleation, annihilation, propagation, and interaction.[21–23] The closed nanoring geometry supports reproducible, non-volatile magnetic configurations, while geometric parameters such as track width and ring overlap, together with the applied magnetic field, determine DW pinning and switching behaviour, thereby shaping the accessible reservoir state space. Nanoring arrays with anisotropic magnetoresistance (AMR) readout have demonstrated strong performance on benchmark reservoir computing tasks, including nonlinear SigT and chaotic time-series prediction, while architectural and field-driven reconfiguration has been shown to further enhance computational performance.[21,22] More recently, Venkat et al. demonstrated that combining multiple nanoring reservoirs further improves task performance, reinforcing the importance of dynamical diversity for computation.[23] Despite these advances, all previous nanoring-based implementations relied on a single spatially averaged AMR output channel, motivating the present work, which exploits width-dependent heterogeneity through multi-channel electrical readout.

In this work, we introduce a nanoring reservoir computer that incorporates controlled geometric heterogeneity within a single array through a systematic gradient in ring track width, directly influencing domain-wall pinning, switching thresholds, and magnetization dynamics. By electrically addressing these subarrays through independent planar Hall effect (PHE) readout channels, multiple width-dependent dynamical responses are accessed in parallel, providing complementary computational features while retaining a single physical reservoir and linear readout architecture. We evaluate the reservoir using nonlinear SigT and MG-TSP tasks at multiple forecast horizons, demonstrating that combining outputs from selected width-dependent channels systematically reduces the normalized root-mean-square error compared with individual-channel readout. We further evaluate task performance as a function of the number of retained principal components, added in order of decreasing variance. The point at which performance saturates provides a task-relevant measure of the reservoir's usable dimensionality, distinguishing genuine dimensionality expansion from redundant or noise-dominated readout channels. Our results demonstrate that engineered geometric heterogeneity transforms a nanoring reservoir from a single-output computational element into a multi-access dynamical building block, providing a pathway towards interconnected magnetic physical computing systems.

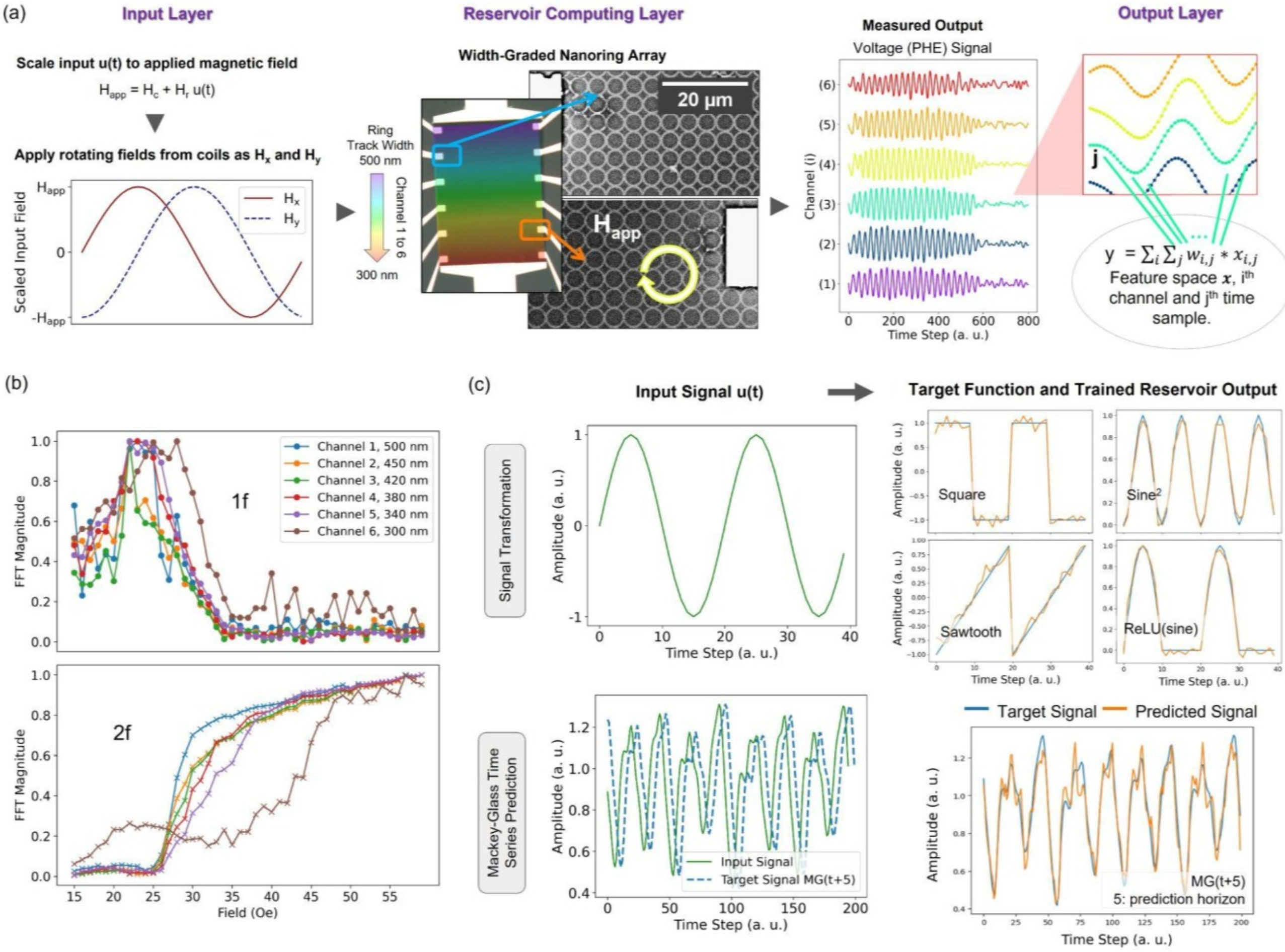

Figure 1. Multi-contact RC pipeline, device behavior, and benchmark tasks. (a) Schematic diagram of the physical RC pipeline: Input signals are mapped onto the amplitude of a rotating magnetic field applied in the plane of the nanoring array. The field excites emergent domain wall dynamics within the array. Output is read via PHE signals between opposite contacts. The reservoir output y is a weighted ($w_{i,j}$) sum of the reservoir states $x_{ij}$, which come from the temporal sampling (j) of the PHE response across each of the six output

channels (i). Microscopy images of the nanoring array are shown in the central panel. (b) Dynamics of the nanoring array: the data shows fast Fourier Transform components measured at the frequency of the driving field (1f) and its 2nd harmonic (2f). Data is shown for each of the 6 PHE output channels. (c) Benchmark tasks performed in this paper. Input signals are shown on the left with an example reservoir output on the right. In the "SigT" task, the input signal is a sine wave (green) and the task is to regress the reservoir output to several other periodic functions (square wave, $sine^2$, sawtooth, ReLU, target = blue, output = orange). In the MG-TSP task, the input signal (green) is the chaotic time series, and the task is to regress the reservoir's output to predict behavior *n* timesteps in the future (target = blue, output = orange). Data is shown for $n = 5$, i.e., approximately a quarter of the signal's quasi-period.

The fabrication, measurement, and computing schemes employed in this work follow the approaches established in our previous studies, reservoir computing with magnetic NRAs.[21,22] The NRAs and electrical contacts are patterned by two-step electron beam lithography with lift-off processing. Metallization of both the $Ni_{80}Fe_{20}$ (10 nm) nanoring arrays and electrical contacts (5 nm Ti/ 100 nm Au) is performed by thermal evaporation. The NRA contains 1024 rings in total. All the rings overlap with their neighbours by 20% of their width and have an outer diameter of 4.4 µm. Each subarray has its own pair of output contacts placed orthogonal to the sense current (~0.5 mA at 13.54 Hz) path, which flows through the large contacts at the top and bottom of the array, such that each channel output provides the PHE signal concentrated on the adjacent subarray. Task-dependent input signals are applied to NRA as in-plane rotating magnetic fields with a set frequency of f = 64 Hz applied by a quadrupole electromagnet. PHE effect measurements are performed by four-point lock-in measurements across all six channels simultaneously.

As demonstrated in our prior work,[22,23] when driven by rotating magnetic fields, NRAs exhibit rich, history-dependent, emergent magnetization dynamics dominated by domain-wall nucleation, propagation, annihilation, and interaction across the arrays. We initially probe these dynamics by examining the PHE signals measured at each contact pair when the NRA is exposed to a variety of constant rotating field amplitudes. Here we have plotted the signal components measured at 1f and 2f, which we have previously shown to represent the stretching of pinned DW and the propagation of free DWs, respectively.[22] The characteristic sigmoidal form of the 2f signal represents the transition from all DWs being pinned to all DWs rotating with the field, with the emergent regime where dynamics are suitable for reservoir computing representing the rising edge of this curve. Notably, this transition occurs at a higher field as the rings track width decreases, due to the modification of domain-wall pinning fields and the relative stability of magnetic microstates. Critically, for a given magnetic field input, each sub-array sits at a different, distinct point within its range of available dynamics such that when the array is subjected to an input sequence where the amplitude of the field is time-varying, there is diversity in the amplitude and temporal structure of the measured voltage responses across the output channels (measured output, Fig. 1(a)). Hence, measuring multiple output channels simultaneously is expected to provide a richer expansion of the input signal than any single channel in isolation.

A schematic diagram of our RC pipeline is illustrated in Fig. 1(c). The NRA acts as a single-input, nonlinear dynamical node driven by the rotating magnetic field. For both tasks studied, the rotating applied magnetic field is defined as $H_{app}(t) = H_0 + H_r u_i(t)$, where $H_0$ is a bias amplitude, and $H_r$ is a scaling field for the input sequence $u_i(t)$. Each input datum $u_i(t)$ thus modulates the applied amplitude field according to $H_{app}(t)$ and is held constant for a single field rotation. Reservoir output is created by temporally sampling the PHE signal to create 50 "virtual nodes" per input channel for each input field cycle, resulting in a total pool of 300 virtual nodes for the device. To explore the effects of multi-channel readout on task performance, we evaluate both individual channels and all possible combinations of *N* channels for $N = 2$ to 6 by concatenating their voltage response to form the reservoir state vector. To suppress readout noise, we bandpass-filter the raw voltage at the 1f and 2f frequencies and average 10 repeats of each experiment run. Reducing per-read noise in this way ensures that any observed performance benefit reflects genuine heterogeneity in the NRA's responses, rather than robustness gained from noise-averaging across many similar (redundant) features. For both tasks, the training, validation, and test sets accounted for 70%, 10%, and 20% of the measured output data, respectively. The reservoir output $y$ is given by $y = \sum_{i,j} w_{i,j} x_{i,j} + b$ where $x_{i,j}$ represents the *j*th temporal sample of the *i*th output channel used in each study. $w_{i,j}$, and $b$ are per-feature weights and a global bias term, respectively, and are trained via ridge regression with the validation set used to find an optimal value of the L2 penalty term. To ensure the repeatability of our results, we create 6 different shuffles of the training, validation, and test sets to calculate the average of the normalized root-mean-square error (NRMSE) with respect to the test data, as well as the standard deviation of the NRMSE as a measure of variability.

We benchmark the system on two tasks: SigT and chaotic MG-TSP (Fig. 1(c)). In the SigT task the input data is ten periods of a sinewave, with $u_i(t)$ sampled at 20 points per period. The reservoir output is trained to transform the sinewave into four nonlinear target functions of the input: square wave, sine squared wave, sawtooth wave, and a ReLU transformation of the input. This task primarily tests the ability of the reservoir to perform non-linear transformations, although memory is also required for waveforms like the sawtooth wave, where identical values of $u_i(t)$ must be mapped to different values of y at different points in the wave's period. In the MG prediction task, $u_i(t)$ is generated from the MG delay differential equation using standard benchmark parameters ($\beta = 0.2$, $\gamma = 0.1$ and $\tau = 17$),[5] and the task is to predict future values of the time series, $MG(t+\Delta t)$, for prediction horizons (future time steps) $\Delta t = 1–20$. Strong performance in this task requires both strong non-linearity and memory from the reservoir. In both tasks, we compare the NRMSE between the reservoir output (orange, Fig. 1(c)) and target (blue, Fig. 1(c)) signals across different channel combinations to assess how access to heterogeneous reservoir dynamics influences computational performance and identify optimal channel subsets.

Results from the SigT task are shown in Figure 2. For each contact combination and values of $H_0$ and $H_r$, we calculate the individual NRMSEs for each target function, as well as the magnitude of a vector of NRMSEs for all four targets, thus measuring the ability of a given contact/field configuration to fit multiple functions. Fig. 2(a) shows a heatmap of the NRMSE vector, annotating in each cell the number of channels, *N*, that needed to be combined to produce the best error for a given ($H_0$, $H_r$) pair. The dark band in the centre of the figure represents the field ranges with best overall performance and is centred on $H_0 \sim 34$ to 30 Oe and $H_r \sim 10$ to 30 Oe. Considering Fig. 1(b) these field scalings will map normalized $u_i(t)$ to values of $H_{app}(t)$ that span most of the non-linear transforms provided by all six channels, thus fully exploiting the device response. Furthermore, the best performance is always provided by 2 - 4 combined channels, illustrating how the heterogeneous response of the device's multiple output channels enhances computation.

To further investigate the effects of combining multiple channels, we explore the NMRSE for all 63 possible channel combinations and each individual waveform target for $H_0 = 36$ Oe and $H_r = 17$ Oe, close to the centre of the band of good performance (Fig. 2(b)). We also highlight the best and worst performing combinations for each value of *N*. We observe that the best performing *N*, as well as the specific channels that must be combined, varies from target to target, although performance is consistently strongest for *N* = 2 - 4 channels. For instance, for the sine-to-square transformation, channels 1, 3, and 5 combined give the best NRMSE, while for the sine to ReLU(sine) task, channels 2 and 5 best accomplish the transformation. Fig. 2(c) shows an example of the best (channels 2 and 4) and worst fit (1 and 6) for the sawtooth waveform where the NRMSE has improved by ~2.6 times. Finally, when we look at the frequency of best performing channels and channel sets (Fig. 2(d)), we observe channels 1 and 2 (left) and *N* = 2 and 3 (right) are leading.

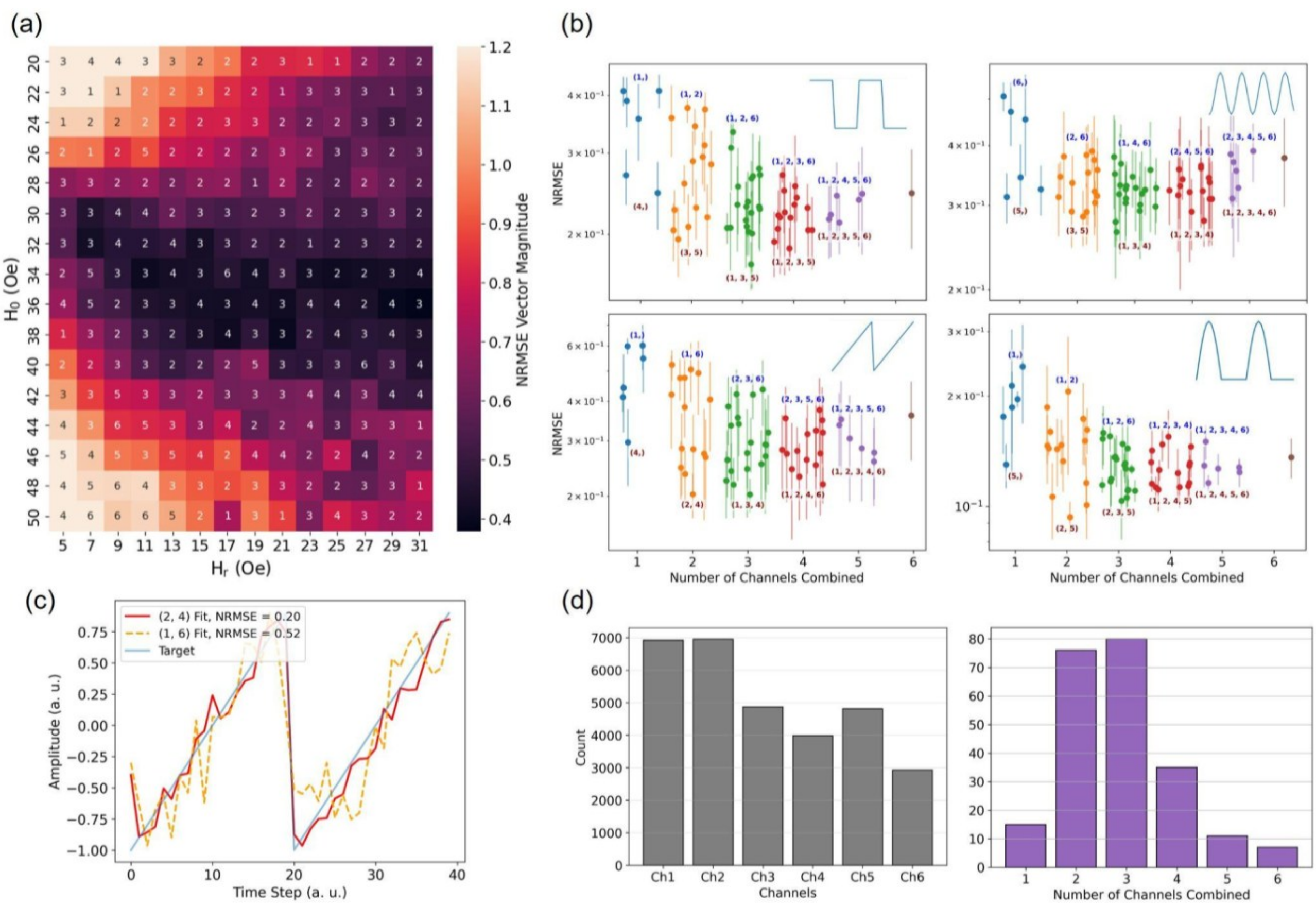

Figure 2. Evaluation of the SigT Task. (a) NRMSE heatmap for the best-performing number of channel combinations for the applied fields, $H_0$ and Hr, respectively. (b) Mapping the NRMSE for all possible channel combinations for each SigT, here $H_0 = 36$ and $H_r$ = 17. The plotted NRMSE points are averaged from 6 different shuffles of the train, validation, and test datasets from the main dataset. The standard deviation is plotted as the error bar of the average point. (c) An example of a good and a bad fit of the sawtooth waveform (target shown) along with the calculated NRMSE for $H_0$ and $H_r$ as (b). (d) Histograms showing the count of the best NRMSE for the channels (left) and the count for the channel combinations (right).

Fig. 3 presents data from the MG-TSP task. Fig. 3(a) presents target and predicted signals for future time steps t +1, t + 5, t + 11, and t + 17, representing a next-step prediction and predictions at ¼, ½, and ¾ signal quasi-periods in the future. In general, signals become harder to predict for points further in the future but are also harder to reconstruct at the ¼ and ¾ points of the quasi-period where the target signal is approximately 90° or 270° out-of-phase with the input, as opposed to in-phase or inverted. Data is shown for the four-channel combination 1, 3, 4 and 5. The NRMSE values at all prediction horizons are shown in Fig. 3(b), which tracks the NRMSE for the $N$ channel combinations with the lowest average prediction error. The double-peaked profile is characteristic of the interactions with the quasi-periodic signal described above. Notably, moving to $N > 1$ is beneficial at all future time-steps, but at the more challenging prediction points the picture is more nuanced with NMRSE values improving up to $N = 5$, and then degrading for $N = 6$ much as in the SigT plots shown in Fig 2(b). However, it is clear from the results reported that optimal combinations are task-dependent, where the best-performing channel combinations differ between SigT and MG-TSP prediction, indicating that combining multiple width-dependent dynamics selectively enhances computation.

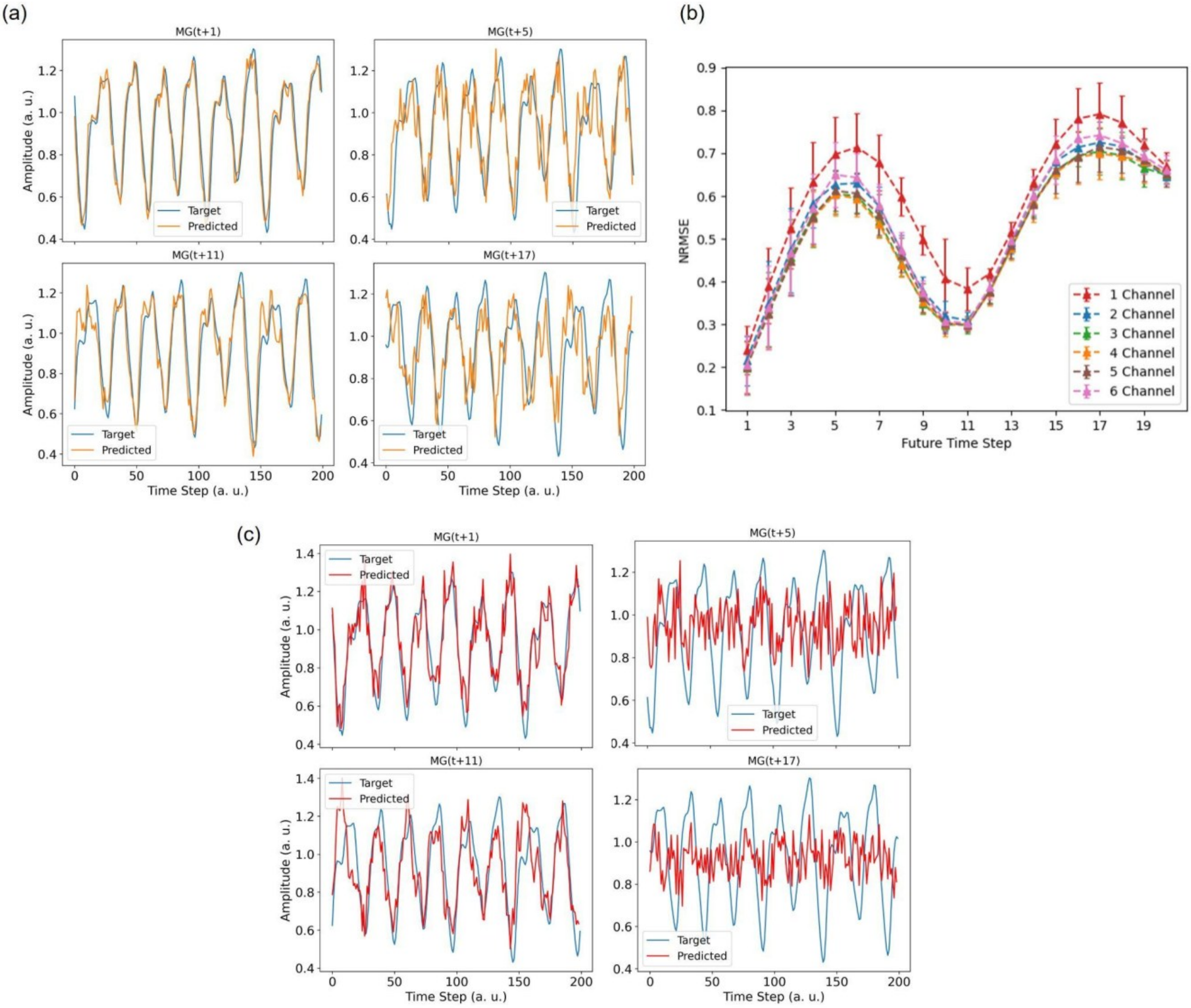

Figure 3. MG-TSP performance. (a) Plots of the predicted and target signals for 1, 5, 11, and 17 prediction horizons for channels (1, 3, 4, 5) as an example. (b) The minimum (best) NRMSE is calculated for different channel set sizes as a function of future time step (prediction horizon); here, $H_0 = 20$ Oe and $H_r = 15$ Oe. (c) Plots of the signal prediction NRMSE case for single channel (here is 6).

One possible interpretation of our results is that the enhancement of task performance at greater $N$ is simply the product of adding more trainable parameters to the model via the larger state vector used. While the degradation of NRMSE values when $N = 6$ in both tasks shows that adding more parameters does not uniformly improve performance, it is still important to establish that this is not the only factor at lower $N$ values and that the improvements in task performance are indeed due to adding more diverse dynamics to the readout.

To address this question, we apply principal component analysis (PCA) on reservoir output for the SigT input sequence (i.e. sinewave). PCA is a standard approach used for dimensionality reduction.[24,25] It simplifies complex, high-dimensional datasets by transforming a large set of variables into a smaller set of uncorrelated variables called principal components (i.e., orthogonal linear combinations of the original variables), while retaining as much of the original data's variance as possible. These components are then ranked based on the amount of variance they capture. We integrate PCA into our readout layer, derived from the combined multi-channel dataset, as the input state space for our ridge regression model. This allows us to answer two questions: (a) How does the performance of different channel numbers and combinations change when using a fixed number of trainable parameters (i.e. by always using the first $k$ PCA components as a readout) and (b) How many genuinely useful readout features does our device possess, and how much of our state space vector is redundant (by exploring how explained variance and task performance vary with $k$).

Fig. 4(a) presents explained variance as a function of $k$ for $N$ = 1 - 6, considering only the channel combinations that achieved the lowest NRMSE on combined waveforms vector magnitude (VM) for each $N$. The data offers two key insights: Firstly, for all values of $N$, total variance is captured within the first 10 principal components across all combinations. This illustrates two key points: Firstly, as all channel combinations saturate the explained variance at relatively low $k$, there are many strongly correlated features in the pool of 50-300 "virtual" nodes measured from the devices. This is not entirely unexpected given the time-multiplexed readout (as temporarily neighbouring "virtual nodes" are likely to be strongly correlated) and the broad similarities of the dynamic responses from the six subarrays (Fig. 1(b)). The data also show that combining multiple outputs genuinely results in a richer state vector, with the number of PCA components required to saturate the explained variance increasing up to $N = 4$.

Fig. 4(b) presents NRMSE VM as a function of the number of principal components used. The results show that we can match the NRMSE of the original full 50-300 output vector datasets (Fig. 2) using a much smaller feature space (<40 PCA components). Furthermore, for $N = 2$-4, progressively more PCA components are required to minimize the NRMSE, with the best NRMSE achieved at $N$ = 3. Collectively, the PCA results demonstrate that computationally useful information is concentrated in a small number of groups of highly correlated features, and that adding more channels to our readout modestly increases the number of these. This substantiates our claim that the heterogeneous dynamics of the NRA create a structural and useful expansion of the reservoir state space.

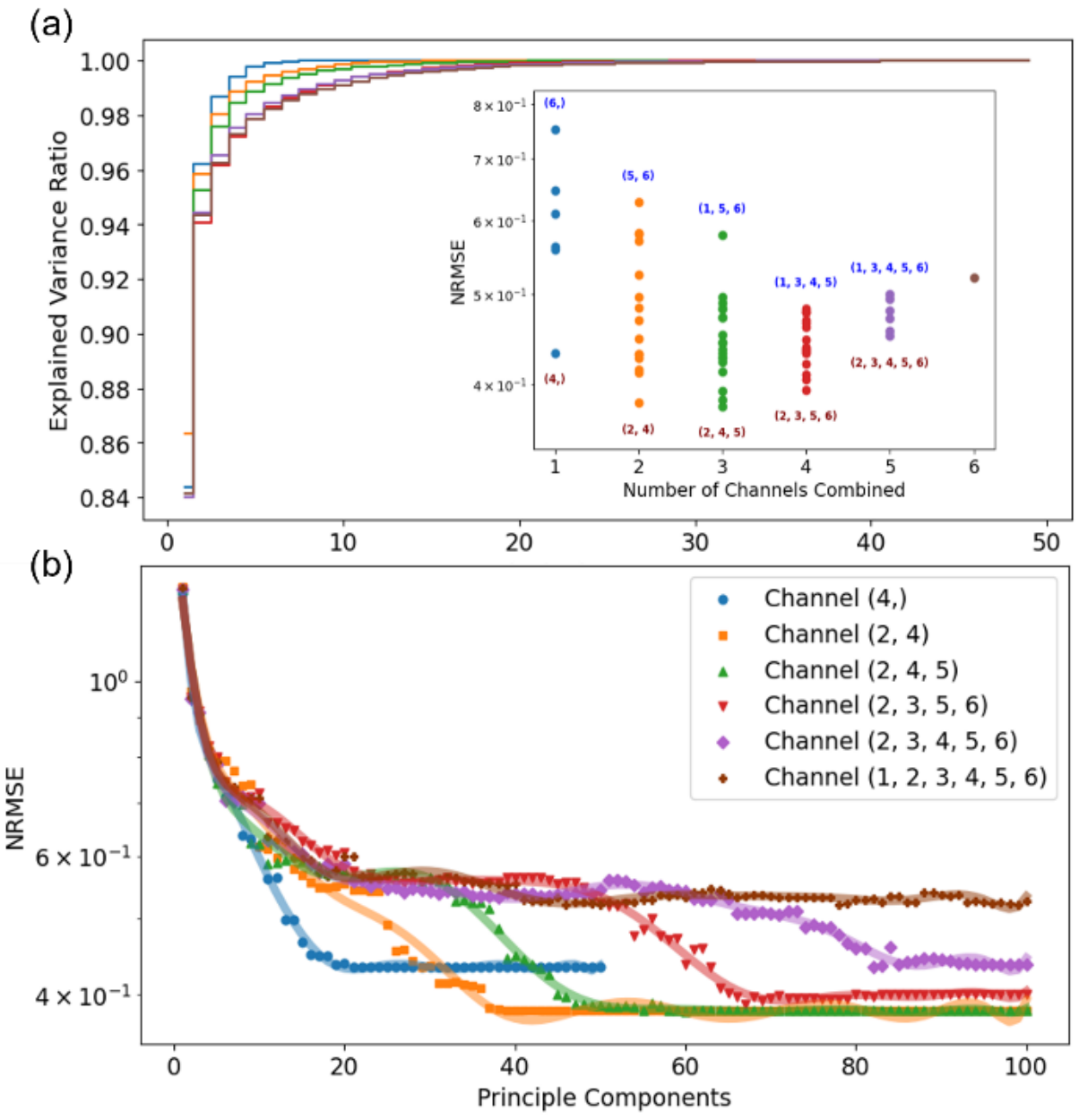

Figure 4. Principal component analysis of physical RC output channel combination for the SigT task of the four waveforms. (a) Cumulative sum of the explained variance ratio of the components calculated for the lowest NRMSE VM channel combination set shown in the inset figure. (b) The NRMSE VM is calculated as a function of the principal components for the different channel combinations. The sinewave input modulated at fields $H_0$ = 36 Oe and $H_r$ = 31 Oe.

In this paper, we experimentally investigated a heterogeneous nanomagnetic reservoir computer based on NRAs with a systematic ring-width gradient (300–500 nm). Independent PHE readout channels were used to access subarrays with distinct widths, enabling the simultaneous measurement of heterogeneous reservoir states and dynamics within a single device. We explored benchmark-task performance in both nonlinear SigT and MG-TSP tasks. In both cases, combining outputs from multiple channels yielded reduced NRMSEs compared to both single-channel and/or all-channel readouts, highlighting the role of geometric heterogeneity in improving reservoir expressivity. PCA analysis was used to establish that these improvements were not purely due to an increase in the number of trainable parameters in the models, and that adding additional channels genuinely created a more diverse feature set to use for computation. Thus, by presenting multiple electrically addressable channels with distinct width-dependent magnetic dynamics, our device provides access to a richer set of dynamical responses and a more expressive reservoir state space.

Reservoir computing constrains expressivity by design, but our results can also be read as a step towards modular building blocks for larger physical computing architectures, an approach we have recently demonstrated in related systems.[26,27] In that framing, heterogeneity buys expressivity that homogeneous devices cannot offer, allowing demanding tasks to be tackled with fewer networked elements. This matters because a principal barrier to scaling such technologies is the difficulty of fabricating and interconnecting large numbers of high-quality devices. Our results therefore point towards magnetic neuromorphic systems in which engineered nanoring reservoirs serve as configurable dynamical nodes, extracting more computation per node rather than demanding more nodes.